# Enhanced ordering in length-polydisperse carbon nanotube solutions at high concentrations as revealed by small angle X-ray scattering

*Vida Jamali[1], Francesca Mirri[1], Evan G. Biggers[1], Robert A. Pinnick,[1] Lucy Liberman[4], Yachin Cohen[4], Yeshayahu Talmon[4], Fred C. MacKintosh[1,2,5,6], Paul van der Schoot[7,8], Matteo Pasquali[1,2,3,9,\*]*

[1]Department of Chemical and Biomolecular Engineering, [2]Department of Chemistry, [3]Smalley-Curl Institute, Rice University, Houston, Texas, 77005, United States.
[4]Department of Chemical Engineering and the Russell Berrie Nanotechnology Institute (RBNI), Technion-Israel Institute of Technology, Haifa 3200003, Israel
[5]Department of Physics and Astronomy, Rice University, Houston, TX 77005, United States
[6]Center for Theoretical Biological Physics, Rice University, Houston, TX 77005, United States
[7]Theory of Polymers and Soft Matter Group, Department of Applied Physics, Eindhoven University of Technology, Postbus 513, 5600 MB Eindhoven, The Netherlands
[8]Institute for Theoretical Physics, Utrecht University, Princetonplein 5, 3584 CC Utrecht, The Netherlands
[9]The Carbon Hub, Rice University, Houston, Texas, 77005, United States.

\*Corresponding author: E-mail mp@rice.edu; Tel +1 713 348 5830.

**Abstract**

Carbon nanotubes (CNTs) are stiff, all-carbon macromolecules with diameters as small as one nanometer and few microns long. Solutions of CNTs in chlorosulfonic acid (CSA) follow the phase behavior of rigid rod polymers interacting via a repulsive potential and display a liquid crystalline phase at sufficiently high concentration. Here, we show that small-angle X-ray scattering and polarized light microscopy data can be combined to characterize quantitatively the morphology of liquid crystalline phases formed in CNT solutions at concentrations from 3 to 6.5 % by volume. We find that upon increasing their concentration, CNTs self-assemble into a liquid crystalline phase with a pleated texture and with a large inter-particle spacing that could be indicative of a transition to higher-order liquid crystalline phases. We explain how thermal undulations of CNTs can enhance their electrostatic repulsion and increase their effective diameter by an order of magnitude. By calculating the critical concentration, where the mean amplitude of undulation of an unconstrained rod becomes comparable to the rod spacing, we find that thermal undulations start to affect steric forces at concentrations as low as the isotropic cloud point in CNT solutions.

**Introduction**

The entropy-driven liquid crystal phase transition in systems of rod-like particles (polymers or colloids) has been a subject of extensive theoretical and experimental research.[1-4] The key factors controlling the phase behavior of such systems are the ratio of the length to the diameter, $L/D$ (the aspect ratio), and the ratio of the persistence length to the length, $P/L$ (the persistence ratio), of the constitutive particles. According to Onsager theory, a solution of infinitely-long rigid rod-like polymers ($L/D \to \infty$, $P/L \to \infty$) will transition from a randomly oriented phase to an orientationally aligned liquid crystal phase at sufficiently high concentrations.[1] Biological systems including tobacco mosaic virus (TMV), *fd* virus, and f-actin are examples of experimental model systems that have been studied to understand the phase behavior of solutions of rod-like particles.[5-8] However, most of these systems are characterized by modest aspect ratios $L/D \approx O(10^2)$, and/or persistence length comparable to the length ($P/L \approx 1$). Therefore, a significant portion of the parameter space remains unexplored. Moreover, these systems are typically charge-stabilized and the electrostatic repulsion is expected to combine with stiffness and aspect ratio to control the type of liquid crystalline phases.[9]

Carbon nanotubes (CNTs) can be synthesized with diameters as narrow as ≈ 0.7 nm and length in excess of 1 µm, leading to aspect ratios of $L/D \approx O(10^3)$. Because of their high bending stiffness, their persistence length ranges from tens to hundreds of micrometers and scales with the cube of the CNT diameter.[10] Therefore, it is possible to synthesize systems with a persistence ratio of $P/L \gg 1$. Aqueous dispersions of CNTs have been achieved by stabilization of the nanotubes using surfactants or DNA.[11-13] However, superacids such as chlorosulfonic acid (CSA) are thermodynamic solvents for CNTs and dissolve CNTs at high concentrations.[14, 15] CSA charges the outer surface of the CNTs, resulting in the electrostatic repulsion, which counteracts the van der Waals attraction between them.[16] The dissolution mechanism is reversible, and, hence, the CNT properties are not affected by the charging process.[15] Therefore, CNTs in CSA are an ideal model system for studying phase behavior of rod-like systems in a previously unexplored range of parameters.

Beyond their application as a model system for fundamental studies of physics of rod-like systems, there is practical interest in studying CNT liquid phases. Out of the various manufacturing methods developed for macroscale CNT materials, solution-processing of CNTs in superacids yields the highest macro scale properties (mechanical strength, thermal and electrical conductivity), while also offering the best promise for scalability.[17-22] Specifically, high-performance CNT fibers are spun out of liquid crystalline CNT solutions.[17] Higher CNT concentration in the liquid crystal phase reduces significantly the superacid solvent use, lowering process costs and $CO_2$ footprint. Therefore, understanding the morphology of CNT liquid crystals at high concentrations is important for the development of these macromaterials.

Earlier studies have shown that solutions of CNTs in CSA transition from an isotropic phase, characterized by randomly oriented rods, to a polydomain nematic phase, where the rods are locally aligned but exhibit no long-range orientational order.[15, 23-25] The transition concentration is inversely proportional to the aspect ratio of the CNTs, following the Onsager prediction corrected for length (aspect ratio) polydispersity.[3, 26] However, it is only recently that there have been studies on the effect of particle diameter and length on the phase behavior of the CNT-CSA liquid crystalline solutions.[27, 28]

Polarized light microscopy is often the simplest method to study the optical texture as a mean to explore the phase diagram of lyotropic liquid crystals. However, it does not provide quantitative information such as rod spacing. Scattering methods accompanied by microscopy techniques are perfectly suited for studying the spatial arrangement of particles in solution, and quantify their liquid crystalline phase behavior. Earlier efforts on using small-angle X-ray scattering (SAXS) focused on studying swollen CNT fibers in acids, and did not study the behavior of solution of individually dispersed CNTs.[29, 30] Recently, small-angle neutron scattering (SANS) was used to probe the effect of CNT length on their phase behavior at concentrations up to 1.5 % by volume.[28] This study showed the formation of a fully nematic phase characterized by a quasi 2D lattice expansion for solutions of long CNTs, while short CNTs exhibited the coexistence of isotropic and nematic phases in the same range of mass concentrations.[28] However, the behavior of this particular system has remained unexplored in higher concentration regimes, where theoretically we expect enhanced ordering.[31]

Here, we use SAXS and polarized light microscopy to study the effect of CNT concentration on the morphology of the liquid crystalline phases formed in CNT-CSA solutions, in the concentration regime that was not explored before. We find that at concentrations as low as 3 % by volume, CNT-CSA solutions form a liquid crystalline phase with pleated optical texture, characterized by a large interparticle spacing. We attribute this large interparticle spacing to electrostatic interactions that couple non-linearly to the CNT bending undulations, increasing tenfold their interaction range. The large interparticle spacing and the scattering signatures could indicate the transition from a nematic to a hexagonally-packed columnar phase, characterized by a six-fold hexagonal symmetry, at the concentrations studied.

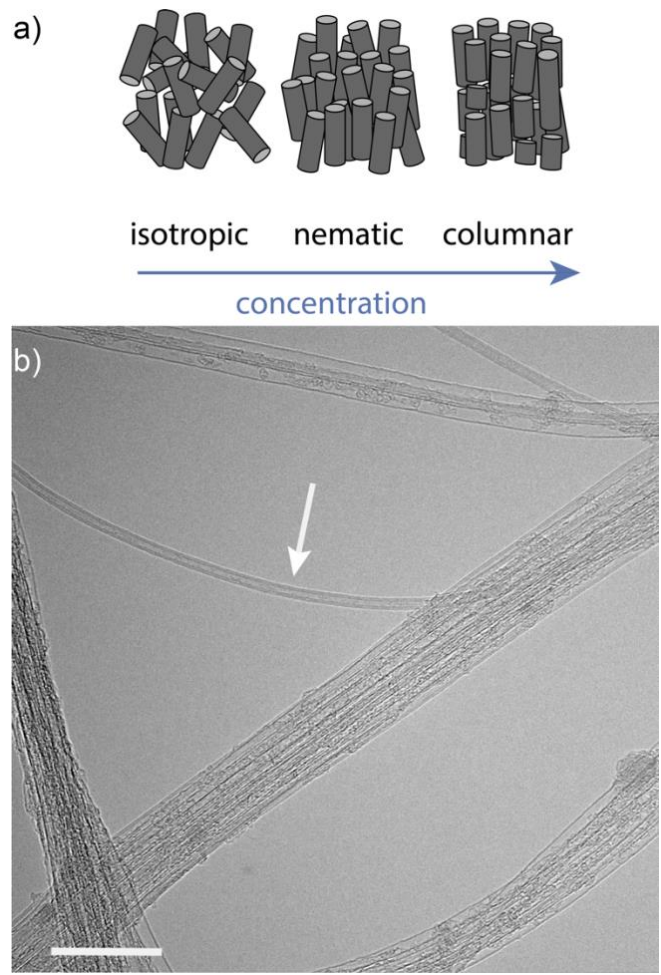

**Figure 1. a)** Schematic representations of isotropic, nematic, and columnar phases formed in length-polydisperse system of rod-like particles. The ratio of the length to the diameter of the rods shown here differs significantly from the case of CNTs ($L/D \approx 5000$), and is only presented for illustrative purposes. **b)** High-resolution transmission electron micrograph of CNTs. The arrow shows an individual CNT of 2.0 nm in diameter. The scale bar corresponds to 20 nm.

**Materials and methods**

*Solution preparation*. CNTs were purchased from Meijo Nano Carbon Co., and purified by thermal annealing at 420 °C for 6 hours in dry air to remove residual amorphous carbon from synthesis. CNTs were then mixed with CSA (Sigma Aldrich, 99 % purity) using a speed mixer (FlackTek, Inc. DAC 150.1speedmixer) for 2.5 hours.

*CNT characterization*. For TEM imaging, a drop (~3 μL) of the CNT-CSA solution was applied onto a perforated carbon film supported on a copper TEM grid (lacey Formvar/carbon films on 200 mesh Cu grids, Ted Pella, Redding, CA, USA), held by tweezers in a controlled environment vitrification system (CEVS).[32] The CEVS was inside a flexible polyethylene "glove-bag" (Sigma-Aldrich), and was kept at 25 °C, while continuously flushed with pure dry nitrogen gas, to prevent moisture penetration. A glass microfiber filter sheet was then used to blot the samples into thin films, followed by plunging them into boiling liquid nitrogen, and quenching into distilled water to coagulate the acid. The grid was then dried overnight in vacuum. The samples were examined with a FEI Talos 200C high-resolution TEM at an accelerating voltage of 200 kV at room temperature. Images were recorded digitally by a 4K×4K pixels, cooled, CMOS-detector FEI Ceta 16M camera, with the TIA software. CNTs diameter $D$ and the number of walls $N$ were estimated from a collection of micrographs obtained (Figure 1b). Additionally, we calculated the density of the CNTs, $\rho_{CNT}$, using the obtained values for the average diameter ($D = 2.04 \pm 0.6$ nm) and the average number of walls ($N = 1.32$), using the expression $\rho_{CNT} = 4000\,[ND - 2\delta_{vdW}\sum_{i=0}^{N-1} i]/(A_s(D+\delta_{vdW})^2)$, which leads to a $\rho_{CNT} \sim 1.3$ g/cm³.[33] Here, $A_s = 1315$ m²/g is the specific area for one side of a graphene sheet, and $\delta_{vdW} = 0.34$ nm is the interlayer van der Waals distance between two CNT walls.[26] The average aspect ratio of the CNTs was estimated using extensional rheology,[26] and found to be $L/D \cong 5000 \pm 100$. The standard deviation is obtained by multiple measurements of the average aspect ratio using the extensional rheology technique. The isotropic-nematic transition point was determined at 0.007 % by volume (70 ppm), using transmitted polarized light microscopy. The degree of length polydispersity $p = \sqrt{\langle L^2 \rangle/\langle L \rangle^2 - 1}$ ($L$ is the length of the constitutive particles and the brackets denote averages) of the CNTs was then estimated as 0.39, using the values for diameter, average length, and the isotropic cloud point, i.e., the concentration at which an infinitesimal amount of nematic phase forms in equilibrium with the isotropic phase.[3, 26] The persistence length of the CNTs, $P$, was estimated as $280 \pm 7$ μm using $P = \pi C D^3/(8K_B T)$, where $C = 345$ J/m² from ab initio calculations of a tube under axial strain.[10]

*SAXS*. Scattering measurements on CNT-CSA solutions were performed using a Rigaku S-Max3000 SAXS machine at the University of Houston, equipped with 3 pinhole collimation. The measurements were carried out in a vacuum chamber, using a 2D detector with $1024 \times 1024$ pixels and a Cu K$\alpha$ X-rays (wavelength of $\lambda_0 = 1.54$ Å) for 24 hours. The instrument was calibrated for $q$ values using Ag behenate standard. The sample-to-detector distance was about 1.143 m and the beam size was 0.2 mm. The samples were prepared by loading the CNT-CSA solutions into rectangular glass capillaries (VitroCom, 1 mm width (larger than the beam size), 0.1 mm path length, and 0.07 mm wall thickness, used as received). A syringe was used to fill the capillaries with high concentration solutions of CNTs in CSA. Rectangular glass capillaries were first attached to 1.5 mm round glass capillary tubes using epoxy glue. The solution was then loaded into rectangular glass capillaries by pulling up the solution, using a syringe inserted into the round capillary tubes. The syringe needle was covered with Teflon tape to ensure a good seal and prevent air from entering the capillary tube. The capillary tubes were subsequently flame-sealed with a butane torch. In order to make the capillary ends more robust in the vacuum chamber, a drop of epoxy was placed on the capillary ends and allowed to dry. The data were calibrated with glassy carbon for intensity, and analyzed using Igor Pro 7 (WaveMetrics, Inc.) and the Irena and Nika packages (available from Argonne National Laboratory).[34, 35] This was done by taking an azimuthal sector average in the direction of maximum scattering intensity, due to the anisotropic scattering pattern arising from the overall alignment of the solutions inside capillary tubes, to obtain intensity as a function of the magnitude of the scattering wave vector $q$, where $q = 4\pi \sin\theta / \lambda_0$, and $2\theta$ is the scattering angle.[34, 35] The accessible range of the wave vector, $q$, is from approximately 0.0075 to 0.25 Å$^{-1}$. The scattering signals were normalized and corrected for background scattering by subtracting the scattering signal of a glass capillary tube filled with solvent. The data was then fitted to a hexagonal close-packed (HCP) cylinder model, using the small-angle diffraction tool in the Irena package, which considers a flat background, a power law scattering ($I(q) \propto q^4$) and three diffraction peaks.[34] Each peak was modeled by a Lorentzian function – see Supporting Information for the details.

*Solution morphology characterization*. Due to the high, broadband light absorption of CNTs, we were not able to image the solutions in the same capillary tubes used for the SAXS measurements. Hence, a drop of CNT solution was sandwiched between two glass slides resulting in a sample thickness of about 10 μm, which was thin enough to allow us to image the sample using transmitted polarized light microscopy. The samples were sealed with Scotch tape to limit the contact between the acid and the moisture in the environment and were imaged immediately after preparation. A Zeiss AxioPlan 2 polarized light microscope equipped with 50x magnification objective, and a cooled charge-coupled device (CCD) camera (AxioCam Zeiss) was used to characterize the birefringent optical texture of CNT liquid crystalline solutions.

**Results and discussions**

We studied five different CNT concentrations ranging from 3 to 6.5 % by volume. Figure 2 shows polarized light micrographs of the as-prepared solutions. The optical texture shows that all samples are birefringent and in the single-phase liquid-crystalline regime. Bright regions are oriented $\pm 45°$ with respect to the cross polarizers of the microscope, indicated by the white arrows. The optical birefringent pattern shows a fan-shaped or pleated ribbon-like texture for the high concentration CNT-CSA liquid crystal solutions. Multiple reasons can give rise to pleated textures in liquid crystal solutions. Shear-induced striations have been reported in the literature for nematic solution of lyotropic and thermotropic liquid crystals[36-38]. Pleated birefringent textures have been also attributed to the higher order hexagonal columnar phases in liquid crystal solutions of length-polydisperse rod-like systems.[39-41] For example, similar microscopic optical textures have been reported for high concentration solutions of poly(γ-octadecyl-L-glutamate) in octadecylamine at 60 % by weight.[39]

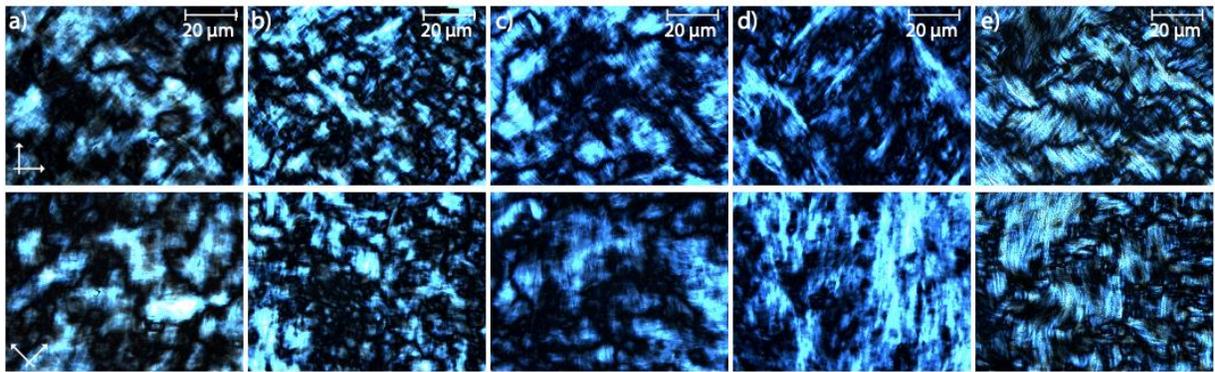

**Figure 2.** Polarized light micrographs showing a pleated texture of solutions of CNTs in CSA at concentrations a) 3, b) 3.8, c) 4.3, d) 5.9, and e) 6.5 % by volume at 0 and 45 degrees with respect to the polarizers of the microscope. Images are post processed to adjust the brightness; the darkest point is set to 0 and the brightest point is set to 255. The crossed arrows show the orientation of the polarizer and analyzer.

To gain better understanding of the effect of concentration on the liquid crystalline behavior of the solutions and the nature of the observed pleated texture, we tested the same solutions using SAXS. The 2D SAXS data (Figure S1) shows an anisotropic pattern due to the overall alignment of the CNT liquid-crystal solutions along the long axis of the capillary tubes. Therefore, the 2D data are sector-averaged to obtain the 1D scattering intensity, $I$, vs. the momentum transfer, $q$, as shown in Figure 3, for the five different concentrations studied.

Upon increasing the concentration, the peaks shift to higher values of $q$, which indicates a tighter CNT packing and an enhanced positional ordering. The relative position of these scattering peaks should indicate the type of liquid crystal phase formed in the solution. Theoretical calculations show that the structure factor for a nematic system of colloidal rods interacting via hard core repulsions presents liquid-like order with multiple peaks, even at relatively low concentrations for scattering vectors along and perpendicular to the director field.[42, 43] Perpendicular to the nematic director, the first peak must be followed by a second peak, spaced at a momentum transfer of a factor of about √5 with respect to that of the first peak–we refer to the Supporting Information for the derivations.[28, 44, 45] Yet, at higher concentration solution studied here, the relative position of the second peak with respect to the first peak for the scattering signals shown in Figure 3 does not follow 1:√5 relative spacing (Figure S3). The second peak is positioned closer to the first peak than what is expected for a nematic liquid crystal phase with a liquid-like order, and is spaced at a distance of √3 relative to the first peak. This relative spacing characterizes the structure of a hexagonal lattice in the plane normal to the director field, i.e., the average axial direction of rods, and could be indicative of the formation of a hexagonal columnar phase, as depicted in the schematic of Figure 1a.

To obtain the positions of the successive correlation peaks, we fitted a hexagonally closed-packed cylinder model to our data for concentrations of 3 to 6.5 % by volume (Figure 3), using the method described in the Materials and Methods section and in the Supporting Information.[34, 35] Table S1 of the Supporting Information shows the fitting parameters and the peak positions for all concentrations studied are presented in Table 1.

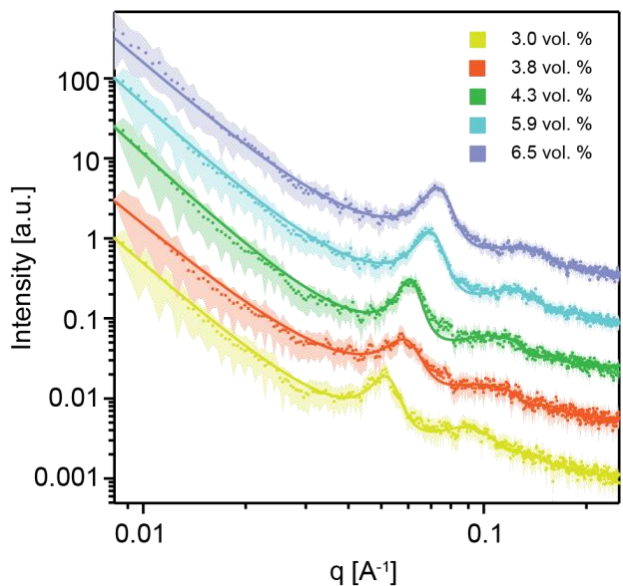

**Figure 3.** SAXS measurements showing the intensity, $I(q)$, vs. momentum transfer, $q$, for CNT-CSA solutions at concentrations ranging from 3 to 6.5 % by volume fitted to a hexagonally close-packed cylinder model ($q_2/q_1 = \sqrt{3}, q_3/q_1 = 2$). The curves are offset vertically for clarity.

**Table 1.** Solution concentration and the position of scattering correlation peaks for CNT-CSA solutions and the corresponding average spacing between neighboring particles, $d$, based on the position of the first peak, $q_1^*$, obtained through curve fitting and using the Bragg's law for the volume fractions ranging from 0.03 to 0.065.

| Volume fraction $\phi$ | Mass Concentration (weight %) | $q_1^*$ (Å$^{-1}$) | $q_2$ (Å$^{-1}$) | Average spacing $d$ (nm) |
|---|---|---|---|---|
| 0.03 | 2.2 | 0.052 | 0.091 | 13.9 |
| 0.038 | 2.8 | 0.060 | 0.104 | 12.1 |
| 0.043 | 3.2 | 0.061 | 0.106 | 11.8 |
| 0.059 | 4.4 | 0.071 | 0.123 | 10.2 |
| 0.065 | 4.8 | 0.075 | 0.129 | 9.6 |

The formation of higher order phases has remained elusive for CNT systems, mainly because high concentration solutions of CNT cannot be obtained by typical dispersion techniques. CNTs typically have a large polydispersity in length, e.g., for CNTs used in this study $p \approx 0.39$. Size polydispersity of particles is known to change the phase behavior of the solution significantly.[46] Upon increasing the concentration, a monodisperse system of rod-like particles is expected to exhibit a transition from a randomly oriented isotropic phase to an orientationally aligned nematic phase, followed by a transition to a smectic phase.[42, 46-48] A smectic phase is characterized by both positional and orientational ordering in the axial direction, where rods are stacked in layers. However, in systems with sufficiently high length polydispersity, stacking of the rod-like particles in the form of layers becomes impossible.[46, 48] In fact, in systems with length polydispersity $p > 0.18$, the smectic phase of hard rods is destabilized completely, leading to a direct nematic-columnar transition at high concentrations.[46] In the columnar phase, the rods are orientationally ordered (similarly to the nematic phase) while having a long-range six-fold hexagonal positional and orientational order in the radial plane. Theoretically, the nematic-columnar phase transition is predicted to occur at volume fractions near 50 %, and is independent of the aspect ratio of the rods.[46] This prediction has been confirmed experimentally for dispersions or suspensions of rod-like particles including viruses, polypeptides, surfactant micelles, and mineral nanorods that show transitions to the columnar phase at concentrations of $\approx 10 - 50$ % by volume.[39, 49-51] The presence of the columnar phase has also been confirmed in DNA dispersions upon evaporating the solvent and increasing the concentration of DNA molecules to $\approx 300$ mg/ml in solutions.[52, 53] More recently, charged imogolite nanotube suspensions have been shown to display nematic-columnar transition at significantly lower concentration of 0.2 % by volume, which was attributed to the large aspect ratio of the nanotubes and to the soft electrostatic repulsion between them.[54] This electrostatic repulsion results in a large Debye screening length ($\approx 30$ nm) that increases the nanotube's effective diameter to $D' \approx D + 2\kappa^{-1}$, where $D'$ is the effective rod diameter, $D$ is the hard-core rod diameter, and $\kappa^{-1}$ is the Debye screening length.[54] This causes their effective volume fraction to be significantly larger than the actual one.[31, 54] A large inter-particle spacing has also been observed in hexagonally assembled suspensions of peptide amphiphile nanofilaments at 0.1 % by volume, where the spacing between the particles have been reported to be 11 times larger than their diameter.[55] The high charge density of the nanofilaments and the resulting electrostatic repulsion between them was identified to be responsible for the large inter-particle spacing in this case as well.

The inter-particle spacing between CNTs in our system can be found using the position of the first correlation peak, $q_1^*$. According to the Bragg's law, for a crystal structure, the characteristic length of the system, $a$, is related to $q_1^*$ through: $a = 2\pi/q_1^*$. This characteristic length scale is directly related to the average spacing between CNTs, $d$. Assuming a hexagonal lattice, $a = d \sin 60°$. For the 6.5 % by volume CNT-CSA solution, the peaks found through the fitted curve are located at $q_1^* = 0.075$, $q_2 = 0.13$, and $q_3 = 0.15$ Å$^{-1}$. This reciprocal spacing translates to a center-to-center distance between neighboring rods $d$ that is equal to about 9.6 nm. Table 1 presents the average spacing between CNTs obtained at different concentrations, assuming a hexagonal lattice for all the concentrations studied.

Figure 4a shows that the center-to-center spacing between neighboring CNTs, $d$, assuming a hexagonal lattice, *vs.* the CNT mass concentration, $c$, which follows the power law of $c^{-1/2}$. The power-law exponent of $-1/2$ suggests a two-dimensional, i.e., radial, lattice expansion of 1D cylinders, which in this case are arranged in a hexagonal lattice. For an aligned system of rods with $L/D \gg 1$, the rods can be assumed as 1D cylinders, where the relative area of the rods with respect to the solution is approximated by their volume fraction.[28] If the rods are packed in a hexagonal lattice, then the average center-to-center spacing between the rods, $d$, is related to the volume fraction $\phi$, through $d = (\pi D^2/(2\sqrt{3}\phi))^{1/2}$.[50, 54, 56, 57] This would allow us to back calculate the diameter of CNTs in solution using the SAXS data, which would verify if the assumption on hexagonal ordering is valid. The average spacing, $d$ is hence plotted vs. $1/\sqrt{\phi}$, as shown in Figure 4b. The diameter, $D$, of CNTs in our system is estimated to be about 2.3 nm from the linear fitting shown in Figure 4b, which is in agreement with the TEM measurements $D = 2.04 \pm 0.6$ nm. Note that the average spacing between the CNTs (e.g., at 3 % by volume, $d = 13.9$ nm) is about 7 times larger than their diameter.

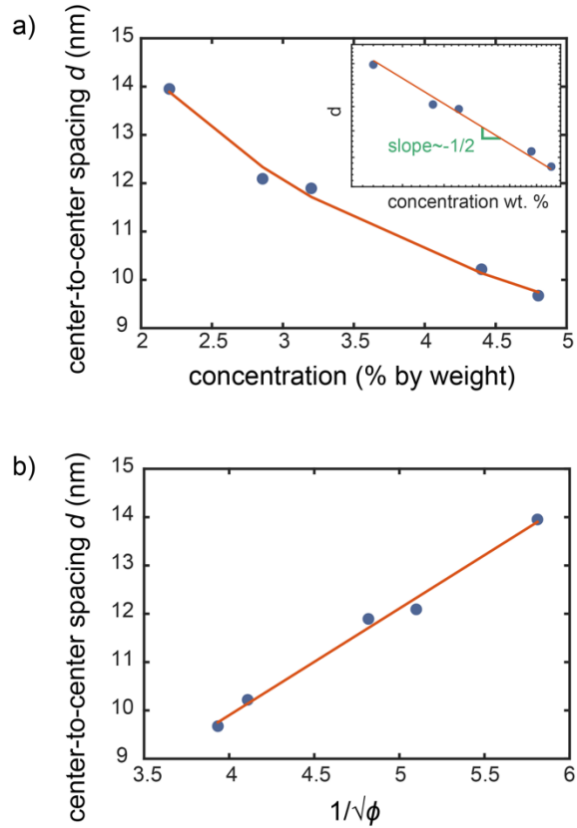

**Figure 4. a) Center-to-center spacing, $d$, between hexagonally packed CNTs in solution (3-6.5% by volume) vs. the CNT mass concentration, and a power-law fit with power-law exponent of -1/2. The inset shows the same plot in logarithmic scale with a fitted line of slope of −1/2 . b) The variation of the spacing $d$, with the corresponding volume fraction of CNTs in solution, $\phi$. The red line represents a linear fit with the slope of 2.2.**

To understand why such a large inter-particle spacing exist in these solutions, we estimated the Debye screening length of the CNTs. In our system, CNTs are positively charged by the acid, resulting in a Debye length of $\kappa^{-1}$, where $\kappa = (e^2 I_s/\epsilon k_B T)^{1/2}$. Here, $e$ is the charge of an electron, $\epsilon$ is the dielectric permittivity, and $I_s$ is the ionic strength of the acid. The ionic strength for an electrolyte of $i$ ionic species of concentration $c_i$ and valence charge $z_i$ is $I_s = \sum_i c_i z_i^2$. For calculating the ionic strength in CNT-CSA system, we consider the ionic contribution from the

CNT protonation by the acid, as well as that of the autoprotolysis of CSA. With CSA being a Brønsted–Lowry acid protonating the sidewalls of CNTs, the protonated CNTs can be considered as positively charged colloidal particles surrounded by $ClSO_3^-$ ions. The fractional charge $\delta^+$ on CNTs in CSA, calculated using the G peak shift of the Raman spectrum[58] (see Figure S3) and Puech's theory,[59] is about 0.024, and is assumed to be independent of the CNT concentration. Hence, the concentration of $ClSO_3^-$ counter ions, $c_{anion}^*$, from CNT protonation is $c_{anion}^* = \phi \rho_{CNT} \delta^+ / M_c$, where $\phi$ is the CNT volume fraction, $\rho_{CNT}$ the density of a CNT, and $M_C$ the molecular weight of carbon. As already alluded to, CSA undergoes autoprotolysis, resulting in $H_2ClSO_3^+$ and $ClSO_3^-$ ions with concentrations $c_{H^+}$ and $c_{anion}$, respectively, which can be calculated using dissociation constant of the acid autoprotolysis: $K_{ap} = c_{H^+} \times c_{anion}$, that is $3 \times 10^{-8}$ mol²/L².[60]

Based on the two ionic contributions mentioned above, the Debye screening length, $\kappa^{-1}$, is about 2 nm for our CNT-CSA system, resulting in the effective diameter $D' = D + 2\kappa^{-1} = 6$ nm. Therefore, the large inter-particle spacing ($d = 11.7$ nm at 4.3 % by volume) in our system cannot be explained solely by the Debye screening length, unlike the case of polyelectrolytes dispersions of peptide amphiphiles and imogolite nanotubes mentioned earlier.[54, 55] In fact, there are multiple terms contributing to the excluded volume between the particles, which are typically neglected. For charged colloidal nanoparticles, large counter ions result in a thick ionic cloud around the nanoparticles, renormalizing the effective volume fraction of the nanoparticles, and leading to stronger electrostatic repulsion.[61] The size of a CSA molecule is about 0.6 nm which is comparable to the diameter of a CNT molecule ($2.04 \pm 0.6$ nm). Most of the classical theories explaining the interaction potentials in colloidal suspensions and electrolytes, including Derjaguin-Landau-Verwey-Overbeek (DLVO) theory, are based on the assumption that the diameter of the colloidal particles is much larger that the solvent and ions surrounding them.[62] In case of nanoparticles, such assumption becomes invalid. Therefore, classical colloidal theories cannot be expected to capture accurately the interaction potential between nanoparticles such as CNTs, as the contribution from the size of the screening ions should become important. Still, even if we include an additional 1.2 nm of spacing between the CNT particles in solution to account for the size of the CSA molecules, we cannot explain the large inter-particle spacing of 11.7 nm obtained from the SAXS results.

A plausible scenario is that the thermal undulations arising from the bending flexibility of CNTs affect the nematic-columnar transition point through increasing the effective diameter of the CNTs.[63, 64] The effect of flexibility of hard rods on the boundaries of the phase diagram in polydisperse systems has been studied extensively in the literature.[65-68] The flexibility of rod-like particles can destabilize the nematic phase in favor of the columnar phase (direct isotropic-columnar transition) in the high flexibility regime ($P/D < 7.745$), while in the low flexibility regime it does not affect the nematic-columnar transition point.[66] Hence, for CNTs, as stiff hard rods ($P/L \gg 1$, and $P/D \gg O(10^3)$), their finite flexibility alone cannot explain the low nematic-columnar transition concentration.[66, 67, 69] Yet, the flexibility of hard rods can drastically enhance the electrostatic repulsion between the particles beyond the Debye screening length explained above by renormalizing the electrostatic potential by a factor of $\frac{1}{4}\exp(\kappa^2\varepsilon^2)$, where $\varepsilon$ is the mean variation of a rod away from its axis.[64, 69] This has been shown to be consistent with experiments on hexagonally-packed DNA-polyelectrolyte gels.[63, 64]

Beyond a critical concentration $\phi^*$, where the interparticle spacing between CNTs becomes comparable to the amplitude of undulations of an unconstrained rod, the rods start to feel each other sterically. For an unconstrained rod ($\phi < \phi^*$), the mean value of the amplitude of undulations about an axis formed by the endpoints is $\varepsilon = ((1/24)(L^3/P))^{1/2}$,[70] which is about 398 nm for CNTs of $L = 10.2$ μm and $P = 280$ μm. This is indeed consistent with previous observations on undulations of individual CNTs in crowded solution using cryo-transmission electron microscopy.[27] By setting the inter-particle spacing equal to the amplitude of the undulations of an unconstrained CNT, we arrive at $\phi^* = 0.0065$ % by volume (65 ppm) as the critical concentration, which is slightly lower than the cloud point of the isotropic phase. Therefore, as soon as these CNTs transition to liquid crystal, $\phi > \phi^*$, and the CNTs are constrained by their neighboring rods, the mean amplitude of the undulations then follows the Odijk's expression that is $\varepsilon = (\lambda^3/P)^{1/2}$, where $\lambda$ is the wavelength of the thermal undulations[*],[64] Such thermal undulations for CNTs as flexible hard rods then result in enhanced electrostatic forces, which increase the effective volume fraction of CNTs in solution, and could explain the large inter-particle spacing that may as well give rise to an early transition to a columnar phase. Future studies could focus on the effect of temperature on CNT thermal undulations and the resulting CNT ordering. Interestingly, higher temperature may shift the transition to higher or lower concentration. On one hand, increasing the temperature lowers the bare persistence length of the CNTs; however, temperature also affects the net charge on the CNTs in a way that is not known at this point. Charge and flexibility couple to how undulations affect in a non-trivial fashion the net repulsion between CNTs and hence the transition concentration from a nematic to a columnar phase.[64] These changes may have further implications for CNT processing into ordered solids such as fibers.

**Conclusion**

---

[*] Note that the nematic interactions modify the tube picture in this case but turn out to only normalize the prefactor and hence are inconsequential to the argument.[67]

We studied the phase behavior of high concentration CNT-CSA solutions by means of polarized light microscopy and SAXS. Our results show that as the concentration of CNTs in solution is increased, the solutions becomes highly ordered with a pleated optical texture. We used SAXS to characterize the interparticle spacing in the high-concentration regimes of liquid crystalline solutions of CNTs in CSA. The presence of high-order peaks confirms the presence of highly ordered phases characterized by a large inter-particle spacing, which is an order of magnitude larger than the diameter of the CNTs in solution. We estimated the effect of steric forces arising from the undulation-enhanced electrostatic repulsions between CNTs. Such repulsive forces are previously known to increase the spacing between semiflexible biopolymers similar to what we found in this study. The increase in the effective diameter results in a higher effective volume fraction, which may give rise to an early transition to a hexagonally packed columnar phase at such low concentrations in this previously unexplored region of parameter space, i.e., aspect ratio and persistence ratio. In addition to its fundamental importance for the phase behavior of rod-like particle systems, this finding in CNT solutions is expected to affect processing, structure, and properties of multifunctional CNT fibers spun from high concentration solutions.

**Acknowledgements**

We thank Ryan Poling-Skutvik, Rana Ashkar, and Lauren W. Taylor for helpful discussions. We thank Dr. Jan Ilavsky for his short course on using Irena and Nika software package for data analysis. Research was supported by Air Force Office of Scientific Research (AFOSR) grants FA9550-12-1-0035 and FA9550-15-1-0370, the Robert A. Welch Foundation (C-1668), the Department of Energy (DOE) award DE-EE0007865 (office of Energy Efficiency and Renewable Energy-Advanced Manufacturing Office), National Science Foundation (NSF) grants CMMI-1025020, DMR-1826623, and PHY-2019745, and United States−Israel Binational Science Foundation (BSF) grants 2012223 and 2016161. The TEM work was performed at the Technion Center for Electron Microscopy of Soft Materials, supported by the Technion Russell Berrie Nanotechnology Institute (RBNI). The SAXS work was performed at the facility of the University of Houston, Department of Chemical and Biomolecular Engineering that is supported by the National Science Foundation under grant DMR-1040446.